\documentclass[journal]{IEEEtran}

\usepackage{amsfonts}
\usepackage{amsmath}
\usepackage{amssymb}
\usepackage{mathabx}
\usepackage{mathrsfs}
\usepackage{tensor}
\usepackage{esint}
\usepackage{tikz}
\usepackage{multirow}
\usepackage{makecell}
\usepackage{textcomp}
\usepackage{amsthm}
\usepackage{subfigure}
\usepackage{graphicx}
\usepackage{stackengine}
\usepackage[caption=false,font=footnotesize,labelfont=sf,textfont=sf]{subfig}
\theoremstyle{remark}
\newtheorem*{observation}{\textbf{Observation}}
\def\delequal{\mathrel{\ensurestackMath{\stackon[1pt]{=}{\scriptstyle\Delta}}}}

\begin{document}
\title{NeRF Solves Undersampled MRI Reconstruction}
\author{\IEEEauthorblockN{Tae Jun Jang$^{1,2}$ and Chang Min Hyun$^{3,*}$}
	\\ \IEEEauthorblockA{$^{1}$School of Mathematics and Computing (Computational Science and Engineering), Yonsei University, Seoul, KR.} \\ \IEEEauthorblockA{$^{2}$AI Vision Group, Samsung Medison, Seoul, KR.} \\ \IEEEauthorblockA{$^{3}$Department of Radiology, Perelman School of Medicine, University of Pennsylvania, Philadelphia, USA.
	\thanks{*Corresponding author (e-mail: chang.minhyun@pennmedicine.upenn.edu).}
	\thanks{The authors declare that they have no known competing financial interests or personal relationships that could have appeared to influence the work reported in this paper.}}}
\markboth{}{ \MakeLowercase{\textit{T. J. Jang and C. M. Hyun}}: }
\IEEEtitleabstractindextext{
\begin{abstract}
This article presents a novel undersampled magnetic resonance imaging (MRI) technique that leverages the concept of Neural Radiance Field (NeRF). With radial undersampling, the corresponding imaging problem can be reformulated into an image modeling task from sparse-view rendered data; therefore, a high dimensional MR image is obtainable from undersampled $k$-space data by taking advantage of implicit neural representation. A multi-layer perceptron, which is designed to output an image intensity from a spatial coordinate, learns the MR physics-driven rendering relation between given measurement data and desired image. Effective undersampling strategies for high-quality neural representation are investigated. The proposed method serves two benefits: (i) The learning is based fully on single undersampled $k$-space data, not a bunch of measured data and target image sets. It can be used potentially for diagnostic MR imaging, such as fetal MRI, where data acquisition is relatively rare or limited against diversity of clinical images while undersampled reconstruction is highly demanded. (ii) A reconstructed MR image is a scan-specific representation highly adaptive to the given $k$-space measurement. Numerous experiments validate the feasibility and capability of the proposed approach.
\end{abstract}
}
\maketitle
\IEEEdisplaynontitleabstractindextext
\IEEEpeerreviewmaketitle

\section{Introduction}
Undersampled magnetic resonance imaging (MRI) has been gaining a great attention to expand our capability of producing cross-sectional MR images with high spatial resolution from optimized data acquisition. It has been particularly anticipated to shorten a long scan time  \cite{haacke1999,sodickson1997,pruessmann1999}, which can contribute to various clinical outcomes such as the increased satisfaction of subjects through a minimized duration time in a uncomfortable narrow bow of MRI machine and the decreased occurrence of motion artifacts induced by deliberate or inevitable movements of subject. Specifically, the undersampled MRI delves into a reconstruction way to minimize time-consuming $k$-space measurements along a phase-encoding direction to the maximum extent possible without compromising the output image quality \cite{hyun2023}.

In standard MRI reconstruction, sampling $k$-space measurements below a certain limit, determined by Nyquist criterion \cite{nyquist1928}, induces image artifacts known as \textit{aliasing} \cite{seo2012}, which can seriously downgrade the quality of resultant MR image. A key of undersampled MRI reconstruction is then how to overcome such artifacts while preserving or, hopefully, restoring image information obtainable from the standard reconstruction with a minimal or more sampling in the sense of Nyquist.

Recently, extensive data-driven approaches with deep learning have been proposed for undersampled MRI reconstruction \cite{chen2022,knoll2020, muckley2021}. These methods have been showing the powerful and promising performance in various tasks such as brain and knee MR imaging. They attempt to learn and take advantage of prior knowledge associated with desired MR images through exploring common inter-training-data relationship under a supervised, unsupervised, or whatever learning framework \cite{hyun2021}. For instance, a paper \cite{hyun2018} leveraged the U-shaped fully convolutional neural network to realize an all-encompassing relation between data distributions in which aliased and corresponding high-quality MR images lie, respectively. A reconstructed MR image is based on the learned group knowledge rather than being highly adaptive to a given measurement. The data-driven approaches seemingly provide the guaranteed effectiveness for samples on or nearby a data distribution similar to a training dataset \cite{hyun2021}. The practical performance is inevitably influenced by the quality, quantity, and diversity of training data.

This study seeks to investigate and suggest a novel approach for data-driven undersampled MRI reconstruction that meets the followings: (i) high performance and robustness regardless of a bunch of training samples for desired MR images and (ii) strong adaptiveness to a specific $k$-space measurement. Thanks to recently emerged Neural Radiance Field (NeRF) techniques \cite{deng2022depth,mildenhall2021,martin2021nerf,yang2022ps}, it becomes enable to accurately represent image rendering in an arbitrary view-point, which implicitly requires to realize an underlying image model, from rendered data in sparse view-points even. With radial undersampling, the corresponding MR imaging problem can be reformulated into an image modeling task from sparse-view rendered data; therefore, a high-quality MR image can be obtained through leveraging the concept of NeRF, namely, implicit neural representation based on a rendering relation. This approach could be a great fit for accomplishing the aforementioned desires.

\begin{figure*}[h]
	\includegraphics[width=1\textwidth]{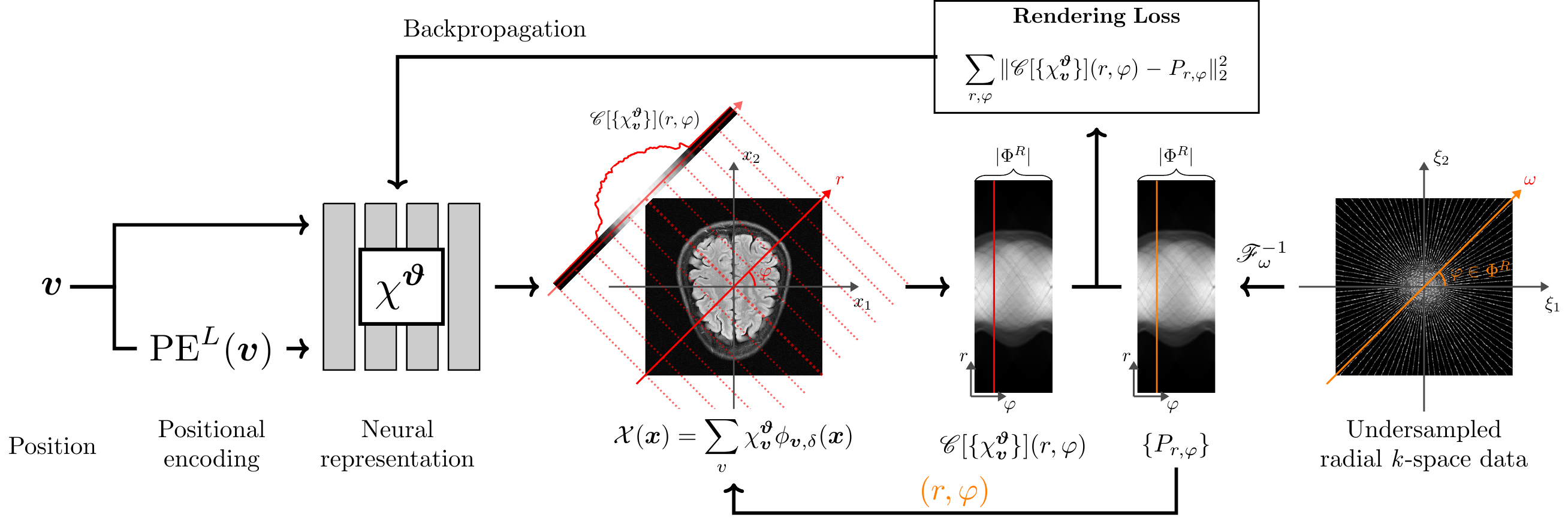}	
	\caption{Implicit neural representation for undersampled MRI reconstruction with radial sampling.}
	\label{main_figure}
\end{figure*}

To be precise, a multi-layer perceptron, designed to output an image intensity from a spatial image coordinate, is learned by minimizing a rendering loss derived from the physical relation between radially undersampled $k$-space data and desired MR image. The learning is based fully on single $k$-space data, not a bunch of measured data and target image sets. The network attempts to realize and take advantage of inter-relation of image intensity over pixels, namely, intra-relation inside a MR image. This is in contrast to existing approaches learning and utilizing inter-relation over MR images. A reconstructed MR image is then a scan-specific representation that is highly adaptive to the specific $k$-space measurement.

Effective undersampling strategies are investigated for high-quality neural representation. We try to provide some rationale somewhat explaining the effective sampling as well as empirical examinations.

Numerous experiments were conducted to validate the feasibility and capability of the proposed reconstruction method by using a publicly open fastMRI dataset \cite{zbontar2018fastmri}. In our empirical experiments, a high-quality MR image could be reconstructed from undersampled $k$-space data even with high acceleration factors. The proposed method is likely to not only improve the image quality but also preserve uncommon anatomical information that tends to be far from shared patterns over MR images. We compared five undersampling strategies: uniform, limited, random, stratified, and golden-angle samplings. The uniform, stratified, and golden-angle tend to provide the better performance than the random and limited under a fixed acceleration factor.

The proposed approach, in fact, is motivated to potentially tackle a challenge in practical diagnostic MR imaging in which data acquisition is relatively rare or limited against  diversity of clinical images while undersampled reconstruction is highly demanded. In fetal MRI, for example, the undersampled reconstruction can be beneficial, since one of major hurdles is fetus-induced motion artifacts because of the long scan time \cite{saleem2014}. Unfortunately, however, existing data-driven methods might not be very powerful as in other MRI applications. It is because fetal MRI is typically utilized as a second-level examination tool followed by ultrasound imaging \cite{manganaro2023}; thereby, gathering fetal MRI data for training may not be flexible in contrast to tremendous image variant factors such as fetal movement, fetal position, gestational age, and etc. In this regard, the proposed method can be viewed as an evolution towards providing an effective solution in this kind of restricted clinical situations, not intending the superiority in all-type imaging environments. 

\section{Method}
For ease of explanation, we restrict ourselves to a 2D MR imaging task. A general 3D reconstruction scheme would be discussed later. Fig. \ref{main_figure} illustrates a schematic description of the proposed method.

\subsection{Inverse problem for undersampled MRI reconstruction}
Let $\mathcal X : \mathbb{R}^2 \rightarrow \mathbb{R}$ be a nuclear spin density distribution of 2D human body to be scanned. We assume that $\mathcal X$ can be expressed as 
\begin{equation} \label{image_assumption}
	\mathcal X (\boldsymbol x) = \sum_{\boldsymbol v \in \mathbb V}X_{\boldsymbol v} \phi_{\boldsymbol v, \delta}(\boldsymbol x) \mathrm{~for~}\boldsymbol x \in \mathbb{R}^2,
\end{equation}
where $\mathbb V$ is a set of indices defined by
\begin{equation}
	\mathbb V \delequal \left\{ (v_1,v_2) \in \mathbb Z \times \mathbb Z : 0 < v_i \leq \dfrac{V}{\delta}, i=1,2 ~\right\},
\end{equation}
and  $\phi_{\boldsymbol v, \delta}$ is a kernel function depending on $\boldsymbol v$ and $\delta$. Here, $\frac{V}{\delta}$ is assumed to be an even number. The distribution of coefficients, $\{ X_{\boldsymbol v} \}$, corresponds to a \textit{MR image}, whose spatial resolution and pixel dimension are associated with $\delta$ and $V$, respectively. One can use a kernel of linear interpolation for $\{X_{\boldsymbol v}\}$, Gaussian kernel, or a square-shaped kernel function given by  
\begin{equation} \label{square-kernel}
	\phi_{\boldsymbol v, \delta}(\boldsymbol x) = \left \lbrace \begin{array}{cl} \dfrac{1}{\delta^2} & \mbox{if } \left|(x_i+\dfrac{V}{2})- (v_i-\dfrac{1}{2})\delta\right| \leq \dfrac{\delta}{2},i=1,2
 \vspace{0.1cm} \\ 0 & \mathrm{otherwise} 
 \end{array}.\right.
\end{equation}

An inverse problem for MRI reconstruction is to recover $\{ X_{\boldsymbol v} \rbrace$ from measured $k$-space data. With radial sampling, a measured dataset can be expressed as $\{K_{\omega,\varphi}\}$, where $K_{\omega,\varphi}$ is given by
\begin{equation} \label{ksample}
		K_{\omega,\varphi} = \mathscr F_{\boldsymbol x}[ \mathcal X ](\omega \Theta_{\varphi}) 
\end{equation}
for some  $\omega \in [-\frac{1}{2\delta},\frac{1}{2\delta}) \mbox{ and } \varphi \in [0,\pi)$. Here, $\mathscr F_{\boldsymbol x}$ is the 2D Fourier transform with respect to $\boldsymbol x$ defined by
\begin{equation}
	\mathscr F_{\boldsymbol x}[f](\boldsymbol \xi) \delequal \int_{\mathbb R^2} f(\boldsymbol x) \exp(-2\pi i \boldsymbol x \cdot \boldsymbol \xi) d\boldsymbol x \mathrm{ ~ for ~ } \boldsymbol \xi \in \mathbb{R}^2,
\end{equation}
$\Theta_{\varphi}=\begin{bmatrix} \cos\varphi, \sin\varphi \end{bmatrix}$, and $\frac{1}{\delta}$ is the minimum interval length for securing spatial information up to the image resolution $\delta$. In standard reconstruction, $\{ K_{\omega,\varphi} \}$ is sampled such that the Nyquist criterion meets, which can be roughly viewed as \cite{seo2012}:
\begin{equation} \label{Nyquist}
	|\{ K_{\omega,\varphi} \}| \gtrapprox |\mathbb V| = \left(\dfrac{V}{\delta} \right)^2,
\end{equation}
where $|\cdot|$ denotes the set cardinality. Specifically, considering a uniform sampling with respect to $\omega$ given by
\begin{equation} \label{radial_uniform}
	\omega \in \left\{- \dfrac{1}{2\delta}, - \dfrac{1}{2\delta} + \dfrac{1}{\sqrt{2}V} , \cdots, -\dfrac{1}{2\delta} + (\dfrac{\sqrt{2}V}{\delta}-1) \dfrac{1}{\sqrt{2}V}\right\},
\end{equation} 
the criterion \eqref{Nyquist} yields the following condition:
\begin{equation} \label{angle_condition}
	N_{\varphi} \gtrapprox \dfrac{V}{\sqrt{2}\delta},
\end{equation}
where $N_{\varphi}$ is the number of radial samples. We here mention that $\frac{1}{\sqrt{2}V}$ in \eqref{radial_uniform} is a sufficient sampling resolution for securing the pixel dimension of $\{X_{\boldsymbol v}\}$, i.e., $\frac{V}{\delta}$.

In this regard, undersampled MRI reconstruction seeks to recover $\{ X_{\boldsymbol v} \}$ while considerably violating the condition \eqref{angle_condition}, i.e., minimizing $N_{\rho}$ as possible. A practical motivation comes from that, in a typical radial sampling set-up, the number of $\varphi$-directional measurements is proportional to a total scan time \cite{feng2022,hyun2020}. 

The associated inverse problem is given by
\begin{equation} \label{IPUMRI_original}
	``\mbox{Recover }\{X_{\boldsymbol v}\} \mbox{ from } \{K_{\omega,\varphi}\}",
\end{equation}
where $\omega$ is given by \eqref{radial_uniform} and $\varphi$ is of $N_{\varphi}$ angles sampled from [0,$\pi$) with $N_{\varphi}<\frac{V}{\sqrt{2}\delta}$. For example, the following uniform undersampling of $\varphi$ is one of options:
\begin{equation} \label{uniform}
	\varphi \in \Phi_{\mathrm{uni}}^{R} \delequal \left\{0, R\pi\left\lfloor\dfrac{\sqrt{2}\delta}{V}\right\rfloor, 2R\pi\left\lfloor\dfrac{\sqrt{2}\delta }{V}\right\rfloor, \cdots\cdots\right\},
\end{equation} 
where $R > 1$ denotes a undersampling (or acceleration) factor and $\lfloor \cdot \rfloor$ denotes the floor operation.

\subsection{Reformulation into an image modeling task from sparse-view rendered data}
We transform $K_{\omega,\varphi}$ into a new data form, $P_{r,\varphi}$, as follows:
\begin{align} \label{projectiontheorem}
	P_{r,\varphi} & \delequal \mathscr F^{-1}_{\omega}[K_{\omega,\varphi}](r),
\end{align}
where $\mathscr F^{-1}_{\omega}$ is the 1D inverse Fourier transform with respect to $\omega$. The Fourier slice theorem provides the following relation between $P_{r,\varphi}$ and $\mathcal X$ \cite{hyun2023_2}: For some $r$ and $\varphi$,
\begin{align} \label{projection}
	P_{r,\varphi} & = \int_{\mathbb R} K_{\omega,\varphi} \exp(2\pi i \omega r) d\omega \nonumber \\
	& = \int_{\mathbb R} (\int_{\mathbb{R}^2} \mathcal X(\boldsymbol x) \exp(-2\pi i \boldsymbol x \cdot \omega \Theta_{\varphi})d\boldsymbol x ) \exp(2\pi i \omega r) d\omega \nonumber \\
	& = \int_{\mathbb{R}^2} \mathcal X(\boldsymbol x) \boldsymbol \delta(\boldsymbol x \cdot \Theta_{\varphi}-r) d\boldsymbol x.
\end{align}
where $\boldsymbol \delta$ is the Dirac delta function. Accordingly, the reconstruction problem \eqref{IPUMRI_original} can be rewritten as 
\begin{equation} \label{IPSVR}
	``\mbox{Recover } \{X_{\boldsymbol v}\} \mbox{ from} ~ \{P_{r,\varphi}\}".
\end{equation}
The reconstruction problem \eqref{IPSVR} can be equivalently viewed as an image modeling task from sparse-view rendered data, which allows us to leverage the idea of NeRF. 

A general image rendering relation comes from the radiative transfer equation given by
\begin{equation} \label{pde}
	\dfrac{dI}{ds} = \epsilon(s) - \tau(s) I(s),
\end{equation}
where $s$ is a length along a ray, $I$ is a ray intensity, $\epsilon$ is a light source, and $\tau$ is an extinction coefficient that can be regarded as the image intensity, i.e., $\tau = \mathcal X$. A solution for \eqref{pde} can be expressed as the following form:
\begin{align}
	I(r,\varphi) = & I_0 \exp(-\int_{\mathbb{R}^2} \mathcal X(\boldsymbol x) \boldsymbol \delta(\boldsymbol x \cdot \Theta_{\varphi}-r) d\boldsymbol x) \nonumber \\ + &\int_{\mathbb{R}} \epsilon_{r,\varphi}(s) \exp(-\int_{s}^{\infty} \mathcal X(r\Theta_{\varphi}+\tilde{s}\Theta_{\varphi}^\perp)d\tilde{s})ds,
\end{align}
where $I_0$ is an initial intensity and $\Theta_{\varphi}^\perp=\begin{bmatrix} \sin\varphi, -\cos\varphi  \end{bmatrix}$. A rendering relation is defined by
\begin{equation} \label{general_relationship}
	\mathscr C[\mathcal X](r,\varphi) \delequal -\log\left(\dfrac{I(r,\varphi)}{I_0}\right).
\end{equation}
Assuming $\epsilon = 0$ (no external source), we then obtain
\begin{equation} \label{rendering}
	\mathscr C[\mathcal X](r,\varphi) = \int_{\mathbb{R}^2} \mathcal X(\boldsymbol x) \boldsymbol \delta(\boldsymbol x \cdot \Theta_{\varphi}-r) d\boldsymbol x = P_{r,\varphi}.
\end{equation}
By reason of \eqref{image_assumption}, we finally have 
\begin{align}
	\mathscr C[\{X_{\boldsymbol v}\}](r,\varphi) = P_{r,\varphi},
\end{align}
where 
\begin{equation} \label{discrete_rendering}
\mathscr C[\{X_{\boldsymbol v}\}](r,\varphi) \delequal \sum_{\boldsymbol v \in \mathbb V} X_{\boldsymbol v} \left (\int_{\mathbb{R}^2} \phi_{\boldsymbol v, \delta}(\boldsymbol x)\boldsymbol \delta(\boldsymbol x \cdot \Theta_{\varphi} - r) d\boldsymbol x \right).
\end{equation}
As a consequence, \eqref{IPSVR} is equivalent to 
\begin{equation} \label{IPSVR_2}
	``\mbox{Recover } \{X_{\boldsymbol v}\} \mbox{ from} ~ \{\mathscr C[\{X_{\boldsymbol v}\}](r,\varphi)\}".
\end{equation}

\subsection{Implicit neural representation for undersampled MRI reconstruction} \label{inr}
To solve \eqref{IPSVR_2}, we define a complex-valued neural network $\chi^{\boldsymbol \vartheta}: \mathbb V \rightarrow \mathbb{C}$, where $\boldsymbol \vartheta$ represents a set of learnable parameters. A network output is given by
\begin{align} \label{MLP}
	\chi^{\boldsymbol \vartheta}_{\boldsymbol v} \delequal \chi^{\boldsymbol \vartheta}(\boldsymbol v) = \mathrm{MLP}^{\boldsymbol \vartheta}(\boldsymbol v,\mathrm{PE}^{L}(\boldsymbol v)),
\end{align}
where $\mathrm{MLP}^{\boldsymbol \vartheta}$ is a multi-layer perceptron that generates two real values in the last layer and then assigns one as a real part and the other as complex, and $\mathrm{PE}^{L}$ is a positional encoding \cite{tancik2020} given by
\begin{equation} \label{positionalencoding}
	\mathrm{PE}^{L}(\boldsymbol v) \delequal \begin{bmatrix} \mathrm{PE}^{L}(v_1),\mathrm{PE}^{L}(v_2)\end{bmatrix}.
\end{equation}
Here, $L$ is some positive constant and $\mathrm{PE}^{L}(v_i)$ is defined by
\begin{align}
	\mathrm{PE}^{L}(v_i) \delequal  [&\cos(2^0\pi v_i), \sin(2^0\pi v_i),\nonumber\\
	&\cos(2^1\pi v_i), \sin(2^1\pi v_i), \nonumber \\ &~~~~~~\vdots~~~~~~~,~~~~~~\vdots \nonumber \\ &\cos(2^{L-1}\pi v_i), \sin(2^{L-1}\pi v_i)] \in \mathbb{R}^{2L}.
\end{align} 
The positional encoding $\mathrm{PE}^{L}$ is a technique that can facilitate to learn high frequency details of pixel intensity distribution \cite{tancik2020}.

The network $\chi^{\boldsymbol \vartheta}$ is trained such that the rendering relation \eqref{rendering} is maximized as follows:
\begin{equation} \label{loss}
	\boldsymbol \vartheta = \underset{\boldsymbol \vartheta}{\mathrm{argmin}} ~ \sum_{\varphi} \sum_{r} ~ \| P_{r,\varphi} - \mathscr C[\{\chi^{\boldsymbol \vartheta}_{\boldsymbol v}\}](r,\varphi) \|^2_2,
\end{equation}
where $\|\cdot \|_2$ represents the Euclidean norm. 

In practice, for ease of computation and training, $\mathscr C[\{\chi^{\boldsymbol \vartheta}_{\boldsymbol v}\}]$ is computed as follows: 
\begin{align} \label{MLPapprox}
\mathscr C[\{\chi^{\boldsymbol \vartheta}_{\boldsymbol v}\}](r,\varphi) & = \sum_{\boldsymbol v \in \mathbb V} \chi^{\boldsymbol \vartheta}_{\boldsymbol v} \left (\int_{\mathbb{R}^2} \phi_{\boldsymbol v, \delta}(\boldsymbol x)\boldsymbol \delta(\boldsymbol x \cdot \Theta_{\varphi} - r) d\boldsymbol x \right) \nonumber \\
	& \approx \sum_{\boldsymbol v \in \mathbb V} \chi^{\boldsymbol \vartheta}_{\boldsymbol v} \left ( \sum_{j} \phi_{\boldsymbol v, \delta}(\boldsymbol x_j) \Delta_{j} \right) \mbox{with } {\boldsymbol x_j \cdot \Theta_{\varphi} = r} \nonumber \\
	& = \left ( \sum_{j} (\sum_{\boldsymbol v \in \mathbb V} \chi^{\boldsymbol \vartheta}_{\boldsymbol v} \phi_{\boldsymbol v, \delta}(\boldsymbol x_j)) \Delta_{j} \right) \nonumber \\
	& \approx \left ( \sum_{j} \chi^{\boldsymbol \vartheta}(\boldsymbol v_j) \Delta_{j} \right ),
\end{align}
where $\boldsymbol x_j$ is a $j$-th position sampled from a line $\boldsymbol x \cdot \Theta_{\varphi} =r$, $\Delta_{j}\delequal \|\boldsymbol x_{j+1} - \boldsymbol x_{j}\|_2$ is the distance between adjacent samples, and $\boldsymbol v_j$ is a relative (real-valued) voxel index corresponding to $\boldsymbol x_j$. The last part of \eqref{MLPapprox} can be understood as a neural network-based approximation for the continuous distribution $\mathcal X$ \cite{montufar2014number}. This can reduce the complexity associated with the kernel function. The reconstruction image is here given by the absolute value of $\{\chi_{\boldsymbol v}\}$.

In addition, $P_{r,\varphi}$ is computed as follows: For a fixed $\varphi$,
\begin{equation}
	P_{r_j,\varphi} \approx (\boldsymbol P_{\varphi})_j \mbox{ and }  \boldsymbol P_{\varphi} \delequal \mathscr F_{\mathrm{disc}}^{-1}[\boldsymbol K_{\varphi}],
\end{equation}
where $\boldsymbol P_\varphi = [P_{r_1,\varphi}, P_{r_2,\varphi}, \cdots ]$, $\boldsymbol K_\varphi = [K_{\omega_1,\varphi}, K_{\omega_2,\varphi}, \cdots  ]$, and $\mathscr F_{\mathrm{disc}}^{-1}$ is the discrete 1D inverse Fourier transform. Here, $r_j$ and $\omega_j$ represent $j$-th samples with respect to $r$ and $\omega$, respectively. Further details for training and network architecture are described in Appendix \ref{TrainingDetails}.

We here mention the reason why the network $\chi^{\boldsymbol \vartheta}$ is designed as complex-valued. In practical MR imaging, various physical and computational factors like noise, magnetic susceptibility, and numerical error can generate complex-valued fluctuations on $P_{r,\varphi}$ in \eqref{loss}.
   
\subsection{Sampling strategy} \label{sampling}
This subsection examines an effective undersampling strategy for high-quality implicit neural representation. The reconstruction problem \eqref{IPSVR} can be equivalently viewed as, for some $R>1$,
\begin{equation} \label{IPUMRI}
	``\mbox{Estimate }\{P_{r,\varphi}\}_{\varphi \in \Phi^1_{\mathrm{uni}}-\Phi^R} \mbox{ from } \{P_{r,\varphi}\}_{\varphi \in \Phi^R}",
\end{equation}
where $\Phi^R \subset \Phi^1_{\mathrm{uni}}$, $R|\Phi^R| \approx |\Phi^1_{\mathrm{uni}}|$, and components of $\Phi^R$ are assumed to be in ascending order. In this perspective, we attempt to discuss the following two questions: (i) what choice of $\Phi^R$ is advantageous to high-quality neural representation and (ii) what benefit is of radial undersampling, compared to others like Cartesian undersampling.

Assuming that $P_{r,\varphi}$ satisfies the Lipschitz property, i.e., for $\varphi \neq \varphi^\prime$,
\begin{equation} \label{Lips}
	\exists ~ M_r>0 ~\mbox{such that}~ |P_{r,\varphi} - P_{r,\varphi^\prime}| \leq M_r |\varphi - \varphi^\prime |,
\end{equation}
the following estimator can be derived: Let $\varphi_j$ denote a $j$-th component of $\Phi^R$. For $\varphi \in \Phi^1_{\mathrm{uni}}-\Phi^R$, 
\begin{align} \label{uniform}
	& |P^{\mathrm{esti}}_{r,\varphi} - P_{r,\varphi}|  \leq \underset{\varphi \in \Phi^1_{\mathrm{uni}}}{\mathrm{max}} |LP_{r,\varphi} - P_{r,\varphi}| = |LP_{r,\varphi_{\star}} - P_{r,\varphi_{\star}}|  \nonumber \\
	& \leq |LP_{r,\varphi_{\star}}-LP_{r,\varphi_{j_{\star}}}| + |P_{r,\varphi_{\star}}-P_{r,\varphi_{j_{\star}}}| \leq C_r|\varphi_{\star} - \varphi_{j_{\star}}| \nonumber \\
	&  \leq C_r \underset{j\in\{1,\cdots,|\Phi^R|-1\}}{\mathrm{max}} ~|\varphi_j - \varphi_{j+1} | \mbox{ for some } C_r>0,
\end{align}
where $j_{\star}$ is an index such that $\varphi_{j_{\star}} \in \Phi^R$ is the nearest point to $\varphi_\star \in \Phi^1_{\mathrm{uni}}$. Here, $P^{\mathrm{esti}}_{r,\varphi}$ is some reasonable approximation for $P_{r,\varphi}$ and $LP_{r,\varphi}$ is a linear interpolation using $\{(\varphi,P_{r,\varphi})\}_{\varphi \in \Phi^R}$. In consequence, we obtain
\begin{equation}
\Phi^R_{\mathrm{uni}} = \underset{\Phi^R}{\mathrm{argmin}} \underset{j\in\{1,\cdots,|\Phi^R|-1\}}{\mathrm{max}} ~ |\varphi_j - \varphi_{j+1} |.
\end{equation}
This explains that undersampling designs with the maximum length between samples as small as possible can be effective. The uniform undersampling $\Phi^R_{\mathrm{uni}}$ is one of good options. 

In turn, we discuss a benefit of radial sampling. If $\mathcal X \in L^2(\mathbb{R}^2)$, $\mathscr F_{\boldsymbol x}\mathcal X$ lies on $L^2(\mathbb{R}^2)$ in the sense of distribution \cite{seo2012}. Here, $L^2(\mathbb{R}^2)$ denotes the Lebesgue space defined by
\begin{equation}
	L^2(\mathbb{R}^2) \delequal  \{ ~ f : \mathbb{R}^2 \rightarrow \mathbb{C} ~ | ~ \| f \|_{L^2} < \infty ~ \},
\end{equation}
where
\begin{equation} \label{eq1}
	\| f \|_{L^2}^2 \delequal \int_{\mathbb{R}^2} |f(\boldsymbol x)|^2 d\boldsymbol x.
\end{equation}
On the other hand, if $\mathcal X \in L^2(\mathbb{R}^2)$, $P_{r,\varphi}$ lies on $H^{t}([0,\pi)\times \mathbb{R})$ for any $0<t\leq\frac{1}{2}$ \cite{epstein2007,lunardi2012} and satisfies \eqref{Lips}. Here, $H^{t}([0,\pi)\times \mathbb{R})$ denotes the fractional Sobolev space defined by
\begin{equation}
	H^{t}([0,\pi)\times \mathbb{R}) \delequal  \{ ~ p: [0,\pi)\times \mathbb{R} \rightarrow \mathbb{C} ~ | ~  \| p \|_{H^{t}} < \infty ~ \},
\end{equation}
where
\begin{align} \label{eq2}
	\| p \|_{H^{t}}^2 \delequal &\int_{0}^{\pi} \int_{\mathbb{R}} |p(r,\varphi)|^2 drd\varphi   \nonumber \\ &+ \int_{0}^{\pi} \iint_{\mathbb{R} \times \mathbb{R}} \dfrac{|p(r,\varphi)-p(r^\prime,\varphi)|^2}{|r-r^\prime |^{1+2t}}  dr^\prime dr d\varphi.
\end{align}
Accordingly, $P_{r,\varphi}$ possesses a higher regularity (smoothness) than $\mathscr F_{\boldsymbol x}\mathcal X$; therefore, the estimation \eqref{IPUMRI} with the high regularity may be more beneficial than general $k$-space estimation with Cartesian or other sampling strategies \cite{bruna2013}.

Experimental examination would be given in Section \ref{subsec:sam}. 

\begin{figure*}[htb]
	\centering
	\subfigure[Reference] {\includegraphics[height=.62\textwidth]{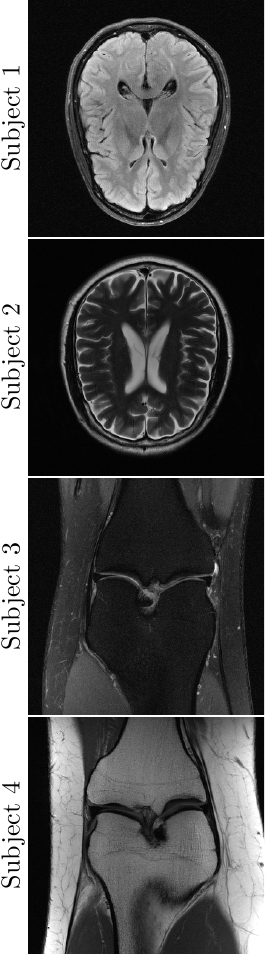}}
		\hspace{-.7em}
	\subfigure[IFFT] {\includegraphics[height=.62\textwidth]{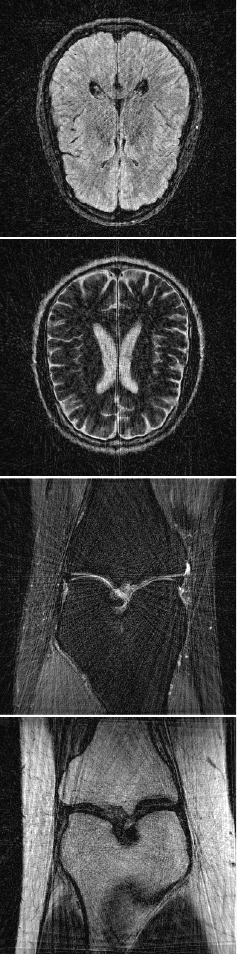}}
		\hspace{-.7em}
	\subfigure[CS] {\includegraphics[height=.62\textwidth]{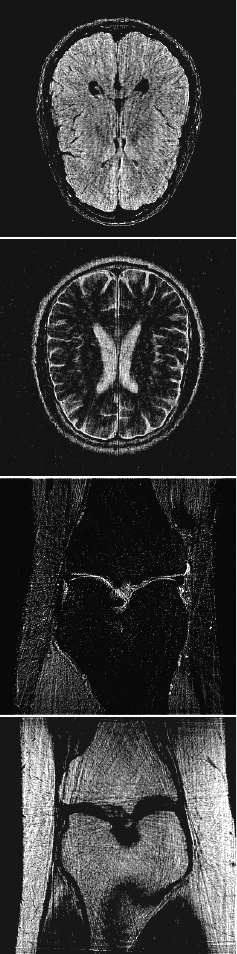}}
		\hspace{-.7em}
	\subfigure[SL] {\includegraphics[height=.62\textwidth]{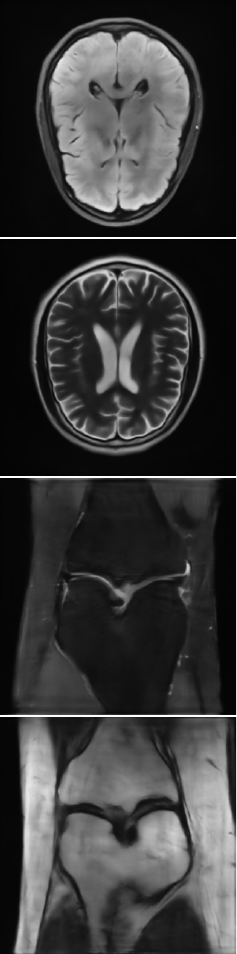}}
		\hspace{-.7em}
	\subfigure[INK] {\includegraphics[height=.62\textwidth]{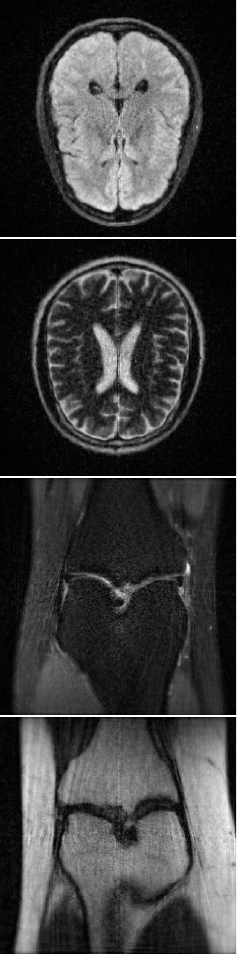}}
		\hspace{-.7em}
	\subfigure[Ours] {\includegraphics[height=.62\textwidth]{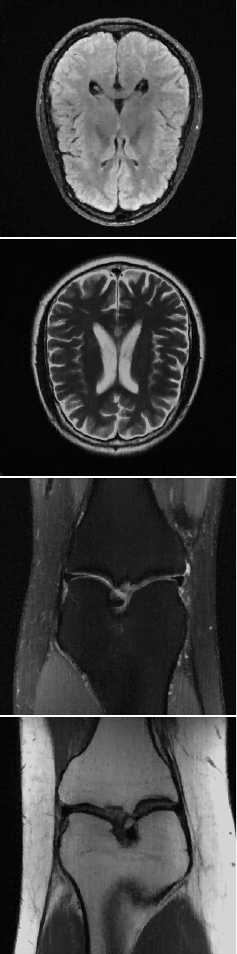}}
	
	\caption{Comparison study: Reconstruction results in four different scans by using IFFT, CS, SL, INK, and our method.}
	\label{fig:result1}
\end{figure*}
\begin{table*}[htb]
	\centering
	\caption{Quantitative evaluation results for comparison study.}
	{ \normalsize
	\begin{tabular}{c||c|c|c|c|c} 
		 {\bf Metric} & {\bf IFFT} & {\bf CS} & {\bf SL} & {\bf INK} & {\bf Ours} \\ \hline \hline 
		\mbox{\bf SSIM($\uparrow$)} & 0.573$\pm$0.069 & 0.521$\pm$0.046 & 0.823$\pm$0.026 & 0.798$\pm$0.052 & \bf 0.904$\pm$0.030 \\ \hline
		\mbox{\bf PSNR($\uparrow$)} & 28.41$\pm$1.395 & 25.59$\pm$1.034 & 27.92$\pm$2.811 & 28.74$\pm$2.667 & \bf 30.16$\pm$2.966 \\ \hline
	\end{tabular}
	}
	\label{table:result1}
\end{table*}

\subsection{Generalization to 3D MR imaging}
This subsection generalizes our reconstruction framework to 3D MR imaging. The following observation provides a base for 3D generalization.

\begin{observation}
	{\it Let $\mathcal X: \mathbb{R}^3 \rightarrow \mathbb{R}$ be a 3D distribution of nuclear spin density. For $w \in \mathbb{R}$, $\varphi \in [0,\pi)$, and $\zeta \in [0,\frac{\pi}{2}]$, we assume that measured 3D $k$-space data is given by
	\begin{equation}
		K_{\omega,\varphi,\zeta} = \mathscr F_{\boldsymbol x}[\mathcal X](\omega \Theta_{\varphi,\zeta}),
	\end{equation}
	where $\Theta_{\varphi,\zeta} = \begin{bmatrix} \cos\varphi\cos\zeta, \sin\varphi\cos\zeta, \sin\zeta \end{bmatrix}$. 
	We define $P$ by 
	\begin{equation}
	P_{r,\varphi,\zeta} \delequal \int_{\mathbb{R}} \mathcal X (\boldsymbol x) \delta(\boldsymbol x \cdot \Theta_{\varphi,\zeta} - r) d\boldsymbol x.
	\end{equation}
	The following relation then holds:
	\begin{equation}
		\mathscr F_{\omega}^{-1}[K_{\omega,\varphi,\zeta}](r)  = P_{r,\varphi,\zeta}.
	\end{equation}}
\end{observation}

The proof can be completed via similar arguments in \eqref{projection} (3D Fourier slice theorem). See \cite{natterer2001mathematics,vassholz2016new} for more details. The observation above provides a rendering relationship for 3D imaging equivalent to \eqref{rendering}. As discussed in Section \ref{inr}, an implicit neural representation can be then learned through minimizing the 3D rendering relation over undersampled data with respect to either $\varphi$ or $\zeta$, or both. Alternatively, the axial extension of the 2D imaging framework can be a simple way. It might be related to practically feasible MR pulse sequence designs.

\section{Result} \label{result_section}
\subsection{Experimental set-up}
In order to evaluate the proposed method, extensive experiments were designed and conducted using a public fastMRI dataset \cite{zbontar2018fastmri} for brain and knee. We obtained 3D $k$-space data and MR volume pairs, whose 2D slice dimension is given by $320 \times 320$. We simulated the radial undersampling acquisition \eqref{ksample} through non-uniform fast Fourier transformation (NUFFT) to a zero-padded image to the size of $452 \times 452$. 

Throughout this section, we followed the convention \cite{zhang2010magnetic} for the number of fully sampled angles (spokes), given by
\begin{equation}
	N_\varphi^{\mathrm{full}}\delequal \left\lfloor \frac{\pi}{2} \times 320 \right\rfloor = 502,
\end{equation}
and for the number of radial directional samples, given by
\begin{equation}
	N_{\omega} \delequal \left\lfloor \sqrt{2} \times 320 \right\rfloor = 452.
\end{equation}
For some radial undersampling with $N_{\varphi}<N_\varphi^{\mathrm{full}}$, we define the acceleration factor $R$ by
\begin{equation}
	R \delequal \dfrac{N_\varphi^{\mathrm{full}}}{N_{\varphi}} > 1.
\end{equation}
For example, $R=10$ means that $N_\varphi=50$ and $N_\omega=452$, i.e., approximate 10 percent spokes were used for reconstruction in terms of $N_{\varphi}^{\mathrm{full}}$. 

For experiments in Section \ref{subsec:comp} and \ref{subsec:acc}, the angles of $N_\varphi$ spokes are determined by the golden-angle sampling \eqref{exp:undersampling} with $R=8$ (62 spokes). In Section \ref{subsec:sam}, we introduce and describe five different sampling schemes including the golden-angle and provide their comparison.

As a quantitative metric, we computed the structural similarity index map (SSIM) and peak signal-to-noise ratio (PSNR), which are defined as follows: For two images $\{\mathcal X_{\boldsymbol v}\}$ and $\{\mathcal Y_{\boldsymbol v}\}$,
\begin{equation}
	\mathrm{SSIM}(\mathcal X,\mathcal Y) \delequal \dfrac{(2m_{\mathcal X}m_{\mathcal Y}+c_1)(2\sigma_{\mathcal X \mathcal Y}+c_2)}{(m_{\mathcal X}^2+m_{\mathcal Y}^2+c_1)(\sigma_{\mathcal X}^2+\sigma_{\mathcal Y}^2+c_2)},
\end{equation}
and
\begin{equation}
	\mathrm{PSNR}(\mathcal X,\mathcal Y) \delequal 10\log_{10}\dfrac{\underset{\boldsymbol v \in \mathcal V}{\mathrm{max}} ~ \mathcal X_{\boldsymbol v}^2}{\displaystyle \sum_{\boldsymbol v \in \mathcal V} (\mathcal X_{\boldsymbol v} - \mathcal Y_{\boldsymbol v})^2},
\end{equation}
where $m_{\mathcal X}$ and $\sigma_{\mathcal X}^2$ represent the mean and variance of $\{\mathcal X_{\boldsymbol v}\}$, $\sigma_{\mathcal X \mathcal Y}$ denotes the covariance between $\{\mathcal X_{\boldsymbol v}\}$ and $\{\mathcal Y_{\boldsymbol v}\}$, and $c_1$ and $c_2$ are positive constants. In this work, all quantitative values were evaluated by taking an average over 30 test data from non-overlapped subjects.

All experiments were conducted in a computer system with two Intel Xeon CPUs E5-2630 v4 and four NVIDIA GeForce GTX 3080ti GPUs.

\subsection{Comparison study} \label{subsec:comp}
This subsection validates the effectiveness of the proposed method and exhibits qualitative and quantitative comparison results to other reconstruction approaches such as inverse fast Fourier transform with zero filling (IFFT), compressed sensing with total variation penalty (CS), supervised learning using TransUNet \cite{chen2021transunet} (SL)  and $k$-space interpolation using neural representation \cite{huang2023neural} (INK).

In Fig. \ref{fig:result1} and Table \ref{table:result1}, qualitative and quantitative comparisons are provided. The proposed method demonstrated the superior performance rather than IFFT, CS, and INK, while SL was comparable. As far as we have implemented, the neural representation in $k$-space appears to be less effective in terms of accuracy and stability. The intensity distribution over pixels in the image domain is strongly correlated even simply in a neighborhood region. However, it is complicatedly entangled in the $k$-space domain and fairly irrelevant in a local region, which might cause the increased training complexity, learning instability, performance degradation, and etc. Compared to SL, the proposed reconstruction tends to keep anatomical details. It appears to be a consequence of the high adaptiveness to the given data. We further elaborate and discuss this in Section \ref{condis}.

Here, CS was implemented using the open-source package, named as \textit{SigPy} \cite{ong2019sigpy}. We trained TransUNet using supervised learning with 532 paired data, whose input is an aliased MR image and label is the corresponding image reconstructed from full sampling. For INK, a multi-layer perceptron \eqref{MLP} was trained, which inputs a image coordinate and outputs a two dimensional vector representing a complex-valued $k$-space intensity. The network was trained by using observed $k$-space data and then used to interpolate unsampled values in $k$-space.

It should be mentioned that the comparison with SL is not fair. We note that the reconstruction approaches other than SL do not require any training data. The use of the larger number of training data can further improve the performance of SL and even can lead it to outperform the proposed method. Most importantly, however, our method is based only on single $k$-space data.

\begin{figure}[t]
	\centering
	\includegraphics[width=.475\textwidth]{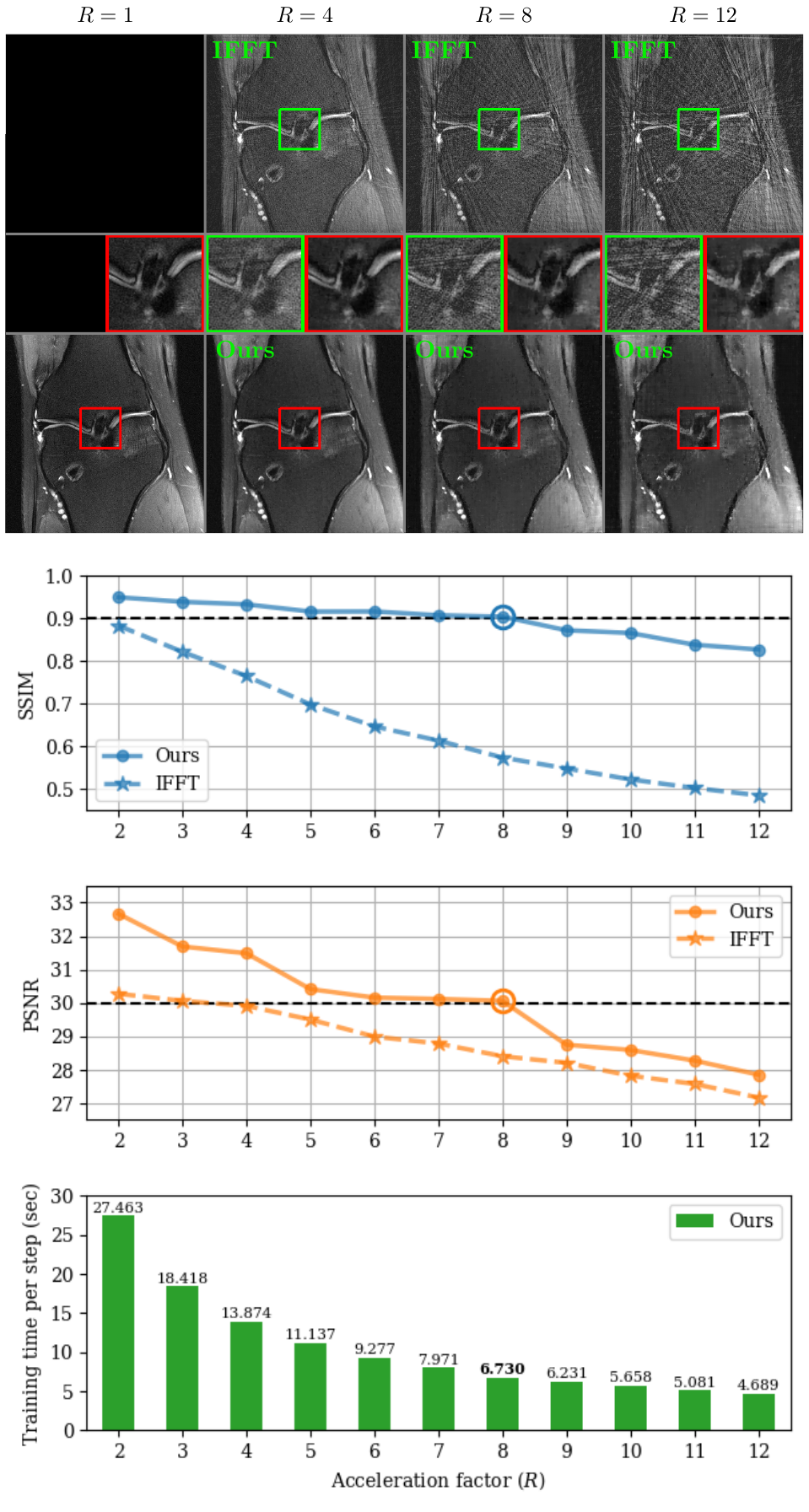}
	\caption{Qualitative and quantitative results for acceleration factor analysis. The top figure shows reconstruction results for IFFT and the proposed method when $R=4,8,12$. Three graphs below present SSIM, PSNR, and training time results.}
	\label{fig:result2}
\end{figure}

\begin{figure*}[t]
	\centering
	\subfigure[Reference]{\includegraphics[width=.153\textwidth]{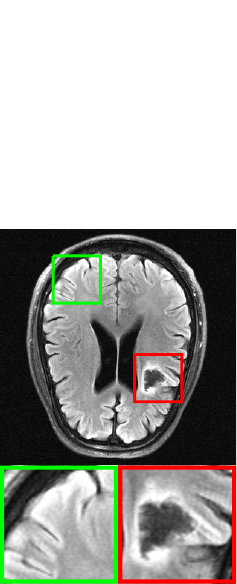}}
	\hspace{-.7em}
	\subfigure[Uniform]{\includegraphics[width=.153\textwidth]{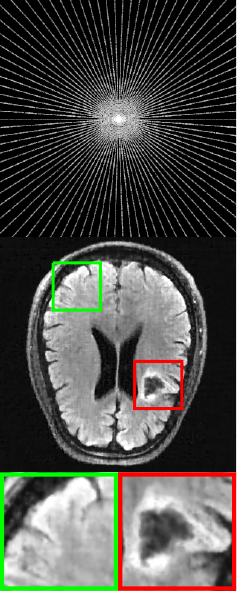}}
	\hspace{-.7em}
	\subfigure[Limited]{\includegraphics[width=.153\textwidth]{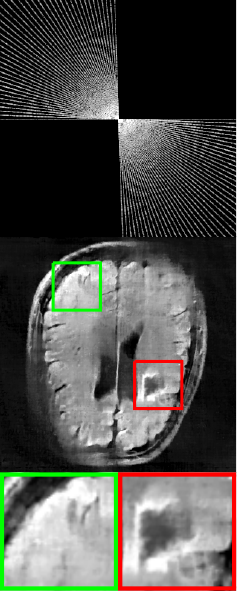}}
	\hspace{-.7em}
	\subfigure[Random]{\includegraphics[width=.153\textwidth]{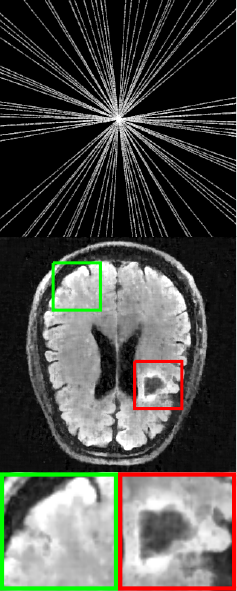}}
	\hspace{-.7em}
	\subfigure[Stratified]{\includegraphics[width=.153\textwidth]{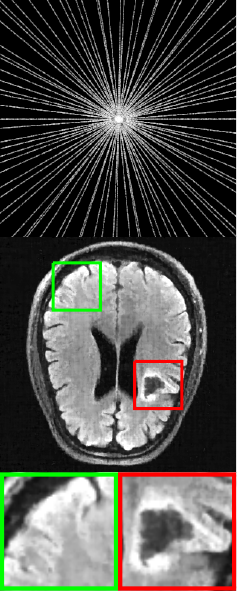}}			
	\hspace{-.7em}
	\subfigure[Golden-angle]{\includegraphics[width=.153\textwidth]{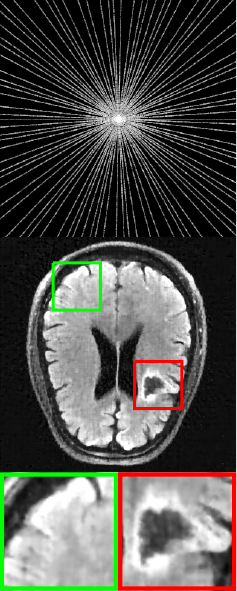}}
	\caption{Qualitative results for sampling study. The top figures in (b)-(f) present radial undersampled $k$-space data for sampling methods. The middle figures show the corresponding reconstruction results by the proposed method. The bottom figures are zoomed images.}
	\label{fig:result3} 
\end{figure*}

\begin{table*}[t]
	\caption{Quantitative evaluation results for sampling study.}
	\centering
	{\normalsize
	\begin{tabular}{c||c|c|c|c|c} 
		 {\bf Metric} & {\bf Uniform} & {\bf Limited} & {\bf Random} & {\bf Stratified} & {\bf Golden-angle} \\ \hline \hline 
		\mbox{\bf SSIM($\uparrow$)} & 0.892$\pm$0.028 & 0.667$\pm$0.084 & 0.875$\pm$0.028 & 0.881$\pm$0.043 & \bf 0.904$\pm$0.030 \\ \hline
		\mbox{\bf PSNR($\uparrow$)} & 28.93$\pm$3.459 & 21.73$\pm$1.967 & 27.49$\pm$2.835 & 29.05$\pm$2.971 & \bf 30.16$\pm$2.966 \\ \hline
	\end{tabular}
	}
	\label{table:result3}
\end{table*}

\subsection{Acceleration factor analysis}\label{subsec:acc}
This subsection examines the effectiveness of the number of spokes to the reconstruction accuracy of the proposed method. Qualitative and quantitative evaluation results are provided in Fig. \ref{fig:result2}. In this experiment, the golden-angle sampling \eqref{exp:undersampling} was used.

We observed reconstruction results by gradually increasing $R$ from 2 to 12. As $R$ is increased (i.e., $N_{\varphi}$ is decreased), the reconstruction ability becomes, of course, weakened. However, the results were constantly superior than those from IFFT in both qualitative and quantitative perspectives. Notably, even for a high acceleration factor of $R=12$ (41 spokes), the proposed method provided a high-quality MR image. Meanwhile, $R$ is associated with the training time of $\text{MLP}^{\boldsymbol \vartheta}$ in \eqref{MLP}, since acquired samples are inputted for training. In other words, the larger $R$, the shorter the training time, as shown in the bottom graph of Fig. \ref{fig:result2}. Note that the inference time is independent to $R$, only affected by the image resolution to be reconstructed. In our case, the inference time was 0.61 seconds approximately for any $R$.

When taking account of the compromise among the reconstruction quality, training time, and scan acceleration, $R=8$ (62 spokes) may be regarded as an empirically optimal choice. 

\subsection{Sampling study}\label{subsec:sam}
This subsection investigates the dependency of the proposed method on undersampling strategies. We here examined the reconstruction performance by varying undersampling schemes. In a fixed acceleration factor of $R=8$ ($N_\varphi=62$), uniform, limited, random, stratified, and golden-angle radial sampling were compared. 

The uniform and limited are to uniformly sample 62 spokes as follows:
\begin{eqnarray}
	&\Phi_{\text{uni}}^{R} &= \left\{0, \frac{\pi}{N_{\varphi}}, \cdots, \frac{(N_{\varphi}-1)\pi}{N_{\varphi}}\right\} \subset [0,\pi), \\
	&\Phi_{\text{lim}}^{R} &= \left\{0, \frac{\pi}{2N_{\varphi}}, \cdots,\frac{(N_{\varphi}-1)\pi}{2N_{\varphi}}\right\} \subset [0,\pi/2).
\end{eqnarray}
We remark that the limited chooses angles within the restricted interval $[0,\frac{\pi}{2})$. The random selects 62 spokes by randomly picking angles in $[0,\pi)$, which can be expressed as
\begin{equation}\label{eq:ransam}
	\Phi_{\text{rand}}^{R} = \{r_1, r_2\cdots,r_{N_{\varphi}}\} \subset [0,\pi).
\end{equation}
The stratified is similar to the uniform sampling, whereas there is perturbation in sampling intervals. It can be expressed as
\begin{equation}
	\Phi_{\text{str}}^{R} = \left\{\frac{r_1}{N_\varphi}, \frac{\pi+r_2}{N_{\varphi}},\cdots, \frac{(N_\varphi-1)\pi+r_{N_\varphi}}{N_{\varphi}}\right\} \subset [0,\pi),
\end{equation}
where each $r_i$ is randomly picked number in $[0,\pi)$ as in \eqref{eq:ransam}.
The golden-angle \cite{feng2022golden} is given by
\begin{equation}\label{exp:undersampling}
	\Phi_{\text{gold}}^{R} = \left\{\text{mod}\left(\frac{(n-1)\pi}{\mathrm{GR}}, \pi\right) : n = 1,\cdots,N_\varphi \right\} \subset [0,\pi),
\end{equation}
where $\mathrm{GR}$ stands for the golden ratio given by $\frac{1+\sqrt{5}}{2}$. 

Fig. \ref{fig:result3} and Table \ref{table:result3} demonstrates qualitative and quantitative comparison results. The golden-angle was the empirical best, complying with reports in conventional radial undersampling MRI \cite{winkelmann2006optimal}. The stratified and uniform were comparable, but less effective than the golden-angle sampling. The limited has the densest spokes between 0 and $\frac{\pi}{2}$, but its result exhibited the lowest performance. The random further improves the reconstruction quality, but it does not still surpass the performance of the uniform, stratified, or golden-angle. This observation agrees with the discussion in Section \ref{sampling}. We note that angle densities in the golden-angle, stratified, and uniform sampling are more even than in the remaining.

\section{Conclusion and Discussion} \label{condis}
This paper seeks to pave a new data-driven imaging way for undersampled MRI reconstruction that can potentially provide an effective solution in clinical applications in which gathering data is restrictive in spite of large variability of target images or tracing anatomical information highly lying on scan or subject is critical. To achieve this, the proposed approach harnesses the power of NeRF that can accurately infer image rendering in an arbitrary view even from sparse-view rendering information. We demonstrated, with radial sampling, the undersampled MR imaging is equivalent to an image modeling from sparse-view rendered data and, therefore, attempted to deal with it by borrowing the idea of NeRF: rendering relation-induced implicit neural representation. One remark is that the representation is, in NeRF, for image rendering implicitly combined with an underlying image model, whereas, in ours, for the direct realization of the model. Numerous experiments validated the feasibility and capability of the proposed method successfully.

We should mention an approach proposed in \cite{shen2022} that does not follow the concept of NeRF but shares the utilization of implicit neural representation. They used a subject-specific priori embedding come from a scan record history, whereas the proposed method is not subjected to the embedding. Since ours follows the formulation of NeRF, in addition, efficient strategies developed along with the advance in the NeRF field can be simply incorporated.

\begin{figure}[t]
	\centering
	\subfigure[Reference]{\includegraphics[width=.12\textwidth]{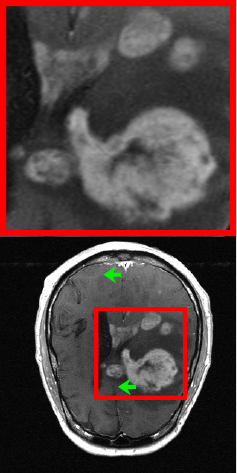}}
	\hspace{-.7em}
	\subfigure[IFFT]{\includegraphics[width=.12\textwidth]{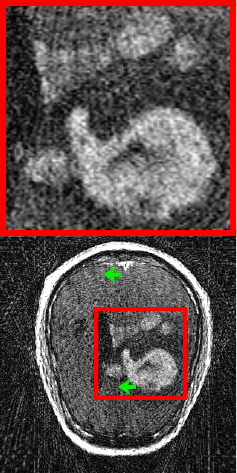}}
	\hspace{-.7em}
	\subfigure[SL]{\includegraphics[width=.12\textwidth]{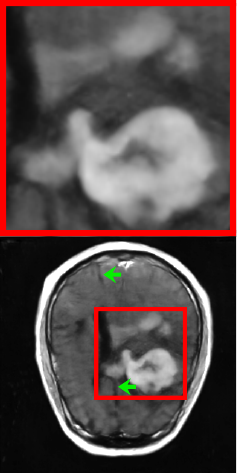}}
	\hspace{-.7em}
	\subfigure[Ours]{\includegraphics[width=.12\textwidth]{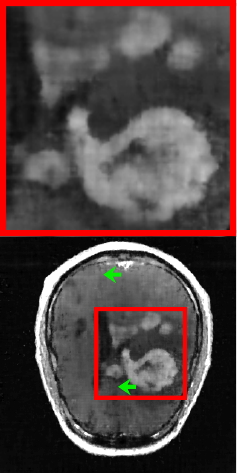}}
	\caption{Results for a case of involved anomalies and high acceleration factor of $R=12$ (41 spokes).}
	\label{fig:discussion1}
\end{figure}

Fig. \ref{fig:discussion1} highlights the advantage of the proposed method. It is an extreme case of involved anomalies and high acceleration factor of $R=12$ (41 spokes). Just only with single radially undersampled $k$-space data, we can obtain a highly improved image in (d), compared to that of the standard IFFT reconstruction with zero-filling in (b). In addition, it is competitive to an image obtained by the TransUnet with supervised leaning of 532 paired data in (c). The proposed approach seems to better capture small variations and uncommon patterns (red boxes) and tends to less produce fake structures (yellow arrows) that do not exist in the reference MR image in (a). It appears to be because the proposed method finds a highly adaptive solution to the given measurement, in contrast, SL attempts to find common patterns over training data and provides an output in some sense of nonlinear averaging  \cite{hyun2021}, possibly resulting in the generation of fake structures and the blurring or distortion of anatomical structures. 

\begin{figure}[t]
	\centering
	\subfigure[reference]{\includegraphics[width=.12\textwidth]{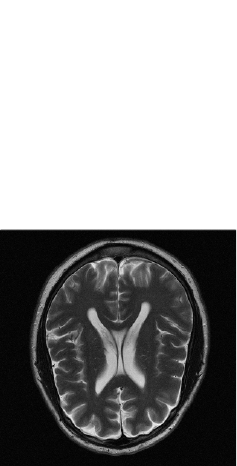}}
	\hspace{-.7em}	
	\subfigure[{$[0,\pi)$}]{\includegraphics[width=.12\textwidth]{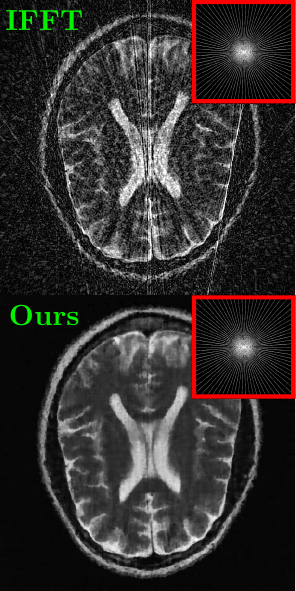}}
	\hspace{-.7em}	
	\subfigure[][{$\left[0,\frac{\pi}{2}\right)$}]{\includegraphics[width=.12\textwidth]{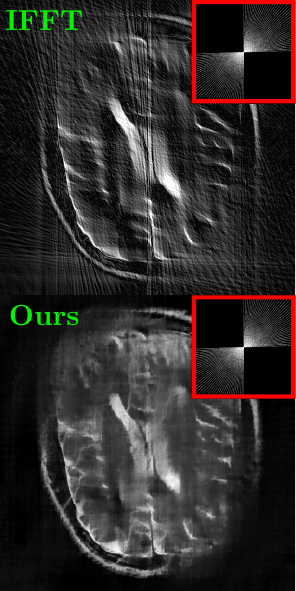}}
	\hspace{-.7em}	
	\subfigure[{$\left[\frac{\pi}{4},\frac{\pi}{2}\right)$}]{\includegraphics[width=.12\textwidth]{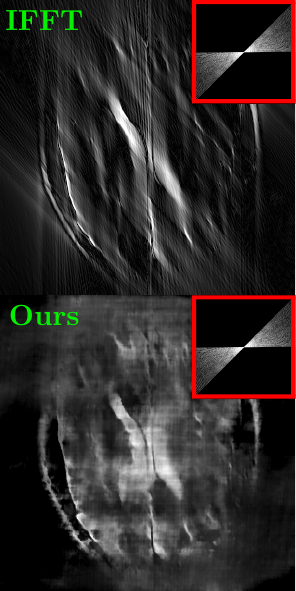}}
	\caption{Results obtained from uniform undersampling in different intervals: (b) $[0,\pi)$, (c) $\left[0,\frac{\pi}{2}\right)$, and (d) $\left[\frac{\pi}{4},\frac{\pi}{2}\right)$. The acceleration factor was $R=12$ (41 spokes).}
	\label{fig:discussion2}
\end{figure}

However, this characteristic can be a double-edged sword. When compared to existing data-driven approaches like SL, our method can be less powerful to produce a plausible (MR image-like) output, as measurement data is less informative in terms of target MR image. Fig. \ref{fig:discussion2} shows this limitation. Under a fixed and high acceleration factor of $R=12$ (41 spokes), three reconstruction results were compared, which were obtained by the proposed method with uniform undersampling in the full angle interval ($[0,\pi)$ in (b)) and limited intervals ($[0,\frac{\pi}{2})$ in (c) and $[\frac{\pi}{4},\frac{\pi}{2})$ in (d)). As the interval range is more limited, that is, measurement data is less informative, the reconstructed image becomes worse. In contrast,   SL can provide a more realistic output as the number of training data and the learning capacity become larger, namely, the better realization on an underlying data distribution over MR images is allowed. The other difficulty is the involved optimization per each reconstruction task. We expect that these aspects might be improved by appropriately integrating both reconstruction ways so that playing complementary roles. This is in our future research direction.

\appendices
\section{Network and Training Details} \label{TrainingDetails}
\begin{figure*}[h]
	\centering
	\includegraphics[width=1\textwidth]{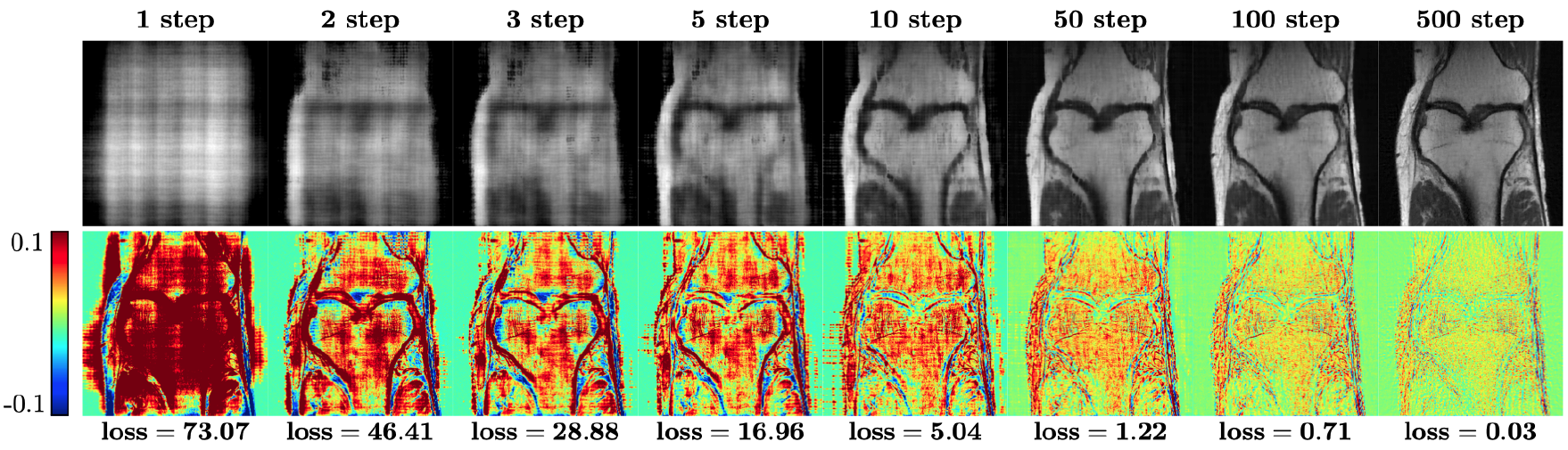}
	\caption{Convergence history of the proposed method when the golden-angle subsampling with $R$=8 (62 spokes) was used.}
	\label{fig:appendix}
\end{figure*}

This appendix provides network and training details. For the implementation of the proposed method, we set the parameter $L$ in the positional encoding in \eqref{positionalencoding} as $L=20$ and used a multi-layer perceptron described in Table \ref{architectures}. For training, we used learning strategies in the vanilla NeRF \cite{mildenhall2021} and Adam optimizer. The number of steps was constantly set as 500. Fig. \ref{fig:appendix} shows a convergence history for one case. 

\begin{table}[h]
	\centering
	\caption{Network architectures}
	{\begin{tabular}{|c|c|c|c|} \hline
		Layer & Input Dim & Output Dim & Activation \\ \hline
		Input & \multicolumn{2}{c|}{82(=2+20$\times$4)} & - \\ \hline 
		Linear & 82 & 256 & sin \\ 
		Linear & 256 & 256 & sin \\
		Linear & 256 & 256 & sin \\	
		Linear & 256 & 256 & sin \\ 
		Concat & 82 & 338 & - \\
		Linear & 338 & 256 & sin \\
		Linear & 256 & 256 & sin \\
		Linear & 256 & 256 & sin \\
		Linear & 256 & 2 & - \\ \hline
		Output & \multicolumn{2}{c|}{2} & - \\ \hline
	\end{tabular}}
	\label{architectures}
\end{table}

\bibliographystyle{IEEEtran} 

\end{document}